\documentclass{article}
\usepackage[hyphens]{url}
\usepackage[utf8]{inputenc}
\usepackage{graphicx}
\usepackage{authblk}
\usepackage[english]{babel}
\usepackage{setspace}
\usepackage{fancyhdr}
\usepackage{array}
\usepackage[margin=1in]{geometry}
\usepackage{float}
\usepackage{amsmath}
\usepackage{amssymb}
\usepackage{bbm}
\usepackage{bm}
\usepackage{scalerel,stackengine}
\usepackage[final]{pdfpages}
\usepackage{longtable}
\usepackage{chngcntr}
\usepackage{placeins}
\usepackage{microtype}
\usepackage{tikz}
\usepackage{makecell}
\usepackage{hyperref}
\usepackage{subcaption}
\usepackage{rotating}
\usepackage{soul}
\usepackage{tabularx} 
\usepackage{pdfpages}
\usepackage{multirow}
\usepackage[square,numbers]{natbib}
\usepackage{listings} 
\definecolor{darkpastelgreen}{rgb}{0.01, 0.75, 0.24}
\definecolor{forestgreen}{rgb}{0.13, 0.55, 0.13} 
\definecolor{limegreen}{rgb}{0.2, 0.8, 0.2}
\definecolor{lavenderindigo}{rgb}{0.58, 0.34, 0.92}
\definecolor{magenta}{rgb}{1.0, 0.0, 1.0}
\definecolor{light-gray}{gray}{0.95}
\hypersetup{
    colorlinks = true,
    citecolor = {blue},
    urlcolor = {blue},
    menucolor = {blue},
    linkcolor = {blue}
}

\makeatletter
\patchcmd{\@maketitle}{\LARGE \@title}{\fontsize{18}{20}\selectfont\textbf{\@title}}{}{}
\makeatother

\title{On the use of non-concurrent controls in platform trials: A scoping review}

\author[1]{Marta Bofill Roig\thanks{marta.bofillroig@meduniwien.ac.at}}
\author[1,2]{Cora Burgwinkel}
\author[3]{Ursula Garczarek}
\author[1]{Franz Koenig}
\author[1]{Martin Posch}
\author[2]{Quynh Nguyen}
\author[2]{Katharina Hees\thanks{katharina.hees@pei.de}}

\affil[1]{Section for Medical Statistics, Center for Medical Statistics, Informatics, and Intelligent Systems, Medical University of Vienna, Vienna}
\affil[2]{Paul-Ehrlich Institut, Department of Biostatistics, Langen} 
\affil[3]{Cytel Inc., Strategic Consulting, Hagen}

\date{}         
\begin{document}

\maketitle

\begin{abstract}
	{\textbf{Background} 
		Platform trials gained popularity during the last few years as they increase flexibility compared to multi-arm trials by allowing new experimental arms entering when the trial already started. Using a shared control group in platform trials increases the trial efficiency compared to separate trials. Because of the later entry of some of the experimental treatment arms, the shared control group includes concurrent and non-concurrent control data. For a given experimental arm, non-concurrent controls refer to patients allocated to the control arm before the arm enters the trial, while concurrent controls refer to control patients that are randomised concurrently to the experimental arm. Using non-concurrent controls can result in bias in the estimate in case of time trends if the appropriate methodology is not used and the assumptions are not met.
		
		\textbf{Methods}  
		We conducted two reviews on the use of non-concurrent controls in platform trials: one on statistical methodology and one on regulatory guidance. We broadened our searches to the use of external and historical control data. We conducted our review on the statistical methodology in 43 articles identified through a systematic search in PubMed and performed a review on regulatory guidance on the use of non-concurrent controls in 37 guidelines published on the EMA and FDA websites. 
		
		\textbf{Results}
		Only 7/43 of the methodological articles and 4/37 guidelines focused on platform trials. With respect to the statistical methodology, in 28/43 articles a Bayesian approach was used to incorporate external/non-concurrent controls while 7/43 used a frequentist approach and 8/43 considered both. The majority of the articles considered a method that downweights the non-concurrent control in favour of concurrent control data (34/43) using for instance meta-analytic or propensity score approaches, and 11/43 considered a modelling-based approach, using regression models to incorporate non-concurrent control data. In regulatory guidelines, the use of non-concurrent control data was considered critical but was deemed acceptable for rare diseases in 12/37 guidelines or was accepted in specific indications (12/37). Non-comparability (30/37) and bias (16/37) were raised most often as the general concerns with non-concurrent controls. Indication specific guidelines were found to be most instructive.
		
		\textbf{Conclusions} 
		Statistical methods for incorporating non-concurrent controls are available in the literature, either by means of methods originally proposed for the incorporation of external controls or non-concurrent controls in platform trials. Methods mainly differ with respect to how the concurrent and non-concurrent data are combined and temporary changes handled. Regulatory guidance for non-concurrent controls in platform trials are currently still limited. }
	{Platform Trial; External controls; Non-concurrent controls.}
\end{abstract}


\newpage

\section*{Background} 

Platform trials aim at evaluating the efficacy of several experimental treatments within a single trial. Experimental arms are allowed to be added or removed as the trial progresses. Furthermore, in platform trials the efficacy compared to control can be tested using a shared control group, which increases the statistical power and reduces the number of required patients as compared to separate trials. Shared controls in platform trials may include concurrent and non-concurrent control data, where, for a given experimental arm, non-concurrent controls refer to data from patients allocated in the control arm before the arm enters the trial. Platform trials have gained popularity during the last years, and there has been much discussion and controversy regarding the use of non-concurrent controls. Non-concurrent controls can further increase the power of the trial, but as the randomisation does not occur simultaneously to treatment arms they can introduce an inflation of the type 1 error rate and bias in the estimates if time trends are present.

The use of randomization in clinical trials has become the gold standard and the proper approach to evaluate a new therapy in a clinical trial is by using a randomized control. However, sometimes the consideration of a randomized control is not feasible as is the case, for instance, in studies for diseases with high mortality or certain rare diseases. In such cases, the use of historical controls has been considered either by substituting the randomized control or by combining them to the randomized control \cite{Collignon2021}. The combination of randomized and historical controls in clinical trials has received much attention over the past several decades, see for example \citep{Jahanshahi2021,Burger2021,Schmidli.2020,Collignon2020,Lim2020} for general discussions on the use of historical controls. Considerable interest has focused on how to combine both controls without introducing bias while reducing the total sample size needed and/or the average total trial duration since Pocock’s seminal paper \cite{Pocock1976} discussed the use of historical controls in the design and analysis of randomised treatment-control trials. For reviews on methods for the inclusion of historical controls, readers may refer to \cite{Viele2014} and \cite{Jiao2019}. 
As with historical controls, non-concurrent controls in platform trials may differ from concurrent controls, and therefore utilising them in the analyses could lead to biases in the estimates if naive analyses are performed \cite{roig2021model}. In the context of historical controls, Viele et al. \cite{Viele2014} 
defined ``drift" as the difference between the true unknown concurrent control parameter and the observed historical control data. To mitigate this problem, several approaches have been proposed for historical controls that could be applied as well to non-concurrent controls in platform trials. The first goal of this paper is to identify the methods currently available for incorporating non-concurrent controls, clarify the key concepts and assumptions, and name the main characteristics of each method.

\begin{figure}[ht]
	\centering
	\includegraphics[scale=0.35]{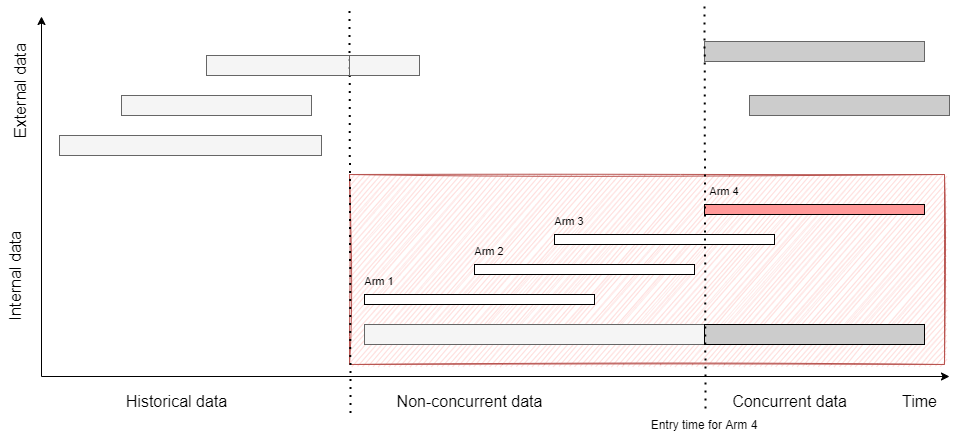}
	\caption{Definition of controls depending on the source and time. The data within the red box represents the (internal) data from a platform trial, the data outside the red box represents the external data. Non-concurrent control data for arm 4 is represented in light grey boxes and concurrent controls are represented in dark grey.}
	\label{fig:defcontrol}
\end{figure}

In the ICH E10 guideline \citep{guideline2000choice}, an externally controlled trial is defined as ``one in which the control group consists of patients who are not part of the randomized study as the group receiving the investigational agent i.e., there is no concurrently randomized control group''. Categorizing them by the time the subject data were collected, Jahanshahi et al. \cite{Jahanshahi2021} distinguish  between the following two types of external controls by: \textit{Concurrent external controls}, as the group of patients recruited to control based on subject level data collected at the same time as the treatment arm but in another setting; and \textit{Non-concurrent external controls} (also referred to as historical controls), as the group based on data collected at a time different (e.g. historical) from the treatment arm (e.g. retrospectively collected from a natural history study, or published data from a previous clinical study). 
Analogously,  internal controls can be defined as the group of patients who are part of the same randomized trial as the group receiving the investigational agent, and divided them as well into: 
\textit{Non-concurrent (internal) controls}: control patients who were recruited before the experimental treatment entered the trial and
\textit{Concurrent (internal) controls}: patients who are recruited to the control when the experimental treatment is part of the trial.  Thus, concurrent control patients have a positive allocation probability of being randomized to the experimental arm. 
In that terminology the non-concurrent controls in platform trials are ``non-concurrent internal controls". Note here that the controls that are most important to distinguish between are: external controls from internal controls; and concurrent internal controls from non-concurrent internal controls. Also note that external control can come from several sources, such as clinical trials, but also "real world data".  
See Figure \ref{fig:defcontrol} for an illustration of different controls definitions. 

Although the inclusion of non-concurrent controls in analyses has been the subject of regulatory discussions for some time, it is still a young topic. Therefore, the second objective of this paper is to summarize the current regulatory view on non-concurrent controls in order to clarify the key concepts and current guidance.  Therefore, we conducted a systematic search in regulatory guidelines regarding the use of non-concurrent controls. 
For the sake of simplicity, we will not use the term ``internal" for concurrent and non-concurrent controls and distinguish them from external controls when it is unclear whether it is concurrent or non-concurrent external controls. Especially methods and regulatory opinions on historical controls and non-concurrent (internal) controls have a considerable overlap, and we will describe both and comment on commonalities and distinctions.
We conclude the  paper with a discussion of the advantages and potential caveats of using non-concurrent controls.  
\section*{Methods}

\subsection*{Review of methods}

\subsubsection*{Search strategy} 
We carried out a systematic search  for the methods in the PubMed database and supplemented the identified with manually searched papers. 
To identify methods  to incorporate non-concurrent control data in PubMed, we performed the following search:
{\small
	\begin{verbatim}
	("non concurrent control*"[Title/Abstract] OR "control arm*"[Title]
	OR "concurrent control*"[Title] OR "historical control*"[Title] 
	OR "external control*"[Title] OR "shared control*"[Title/Abstract]) 
	AND 
	("design*"[Title/Abstract] OR "stud*"[Title/Abstract] 
	OR "platform trial*"[Title/Abstract] 
	OR "master protocol*"[Title/Abstract]) 
	AND 
	("trial*"[Title/Abstract] OR "clinical"[Title/Abstract])
	\end{verbatim}}
This study is reported in accordance with the PRISMA reporting guideline's extension for scoping reviews \cite{Pagen71}. 
Other options including the term "historical controls" were taken into consideration. However, such searches resulted in a large number of articles (for instance, including the term "historical control*" in the search as [Title/Abstract] returns 3209 articles) that were unfeasible to review and most of which were not of interest for this review. Therefore, it was decided to perform a narrower search, prioritising the relevance of the articles in terms of the methodology, and checking that the  predefined a priori list of articles of interest was included and that the most relevant articles reviewing historical controls methods were also included. 

\subsubsection*{Identification of articles}

In order to be included for data extraction, the focus of the article had to be one of the following: 
1)	The description of a method proposed to include external/non-concurrent controls and concurrent controls in clinical trials. 
2)	The article considered an application of a method which included external/ non-concurrent controls together with concurrent controls in a clinical trial context with a detailed description of the method used. 
3)	The article is an overview of several methods to include external controls (e.g. review article). 
4)	The article is about the comparison of several methods (e.g. via simulation studies) to include external controls.
On the other hand, the article was not considered if: 
1)	The article focused on shared concurrent controls but not on the inclusion of external/non-concurrent controls; 2) The term external/non-concurrent control was used in another context; 
3)	The article focused on a clinical trial using external/non-concurrent controls, but not on the methods.

The screening was performed by two reviewers in a three-step process: 
1)	Article titles were screened by a reviewer and selected based on the inclusion criteria or excluded based on the exclusion criteria.  
2)	Article abstracts were screened and selected by a reviewer based on the inclusion criteria or excluded based on the exclusion criteria.  
3)	Selected articles were fully read and selected by two reviewers based on the inclusion criteria or excluded based on the exclusion criteria.  
If the selected articles referred to other relevant articles not found by our search, these were added retrospectively. 

\begin{figure}[ht]
	\centering
	\includegraphics[scale=0.4]{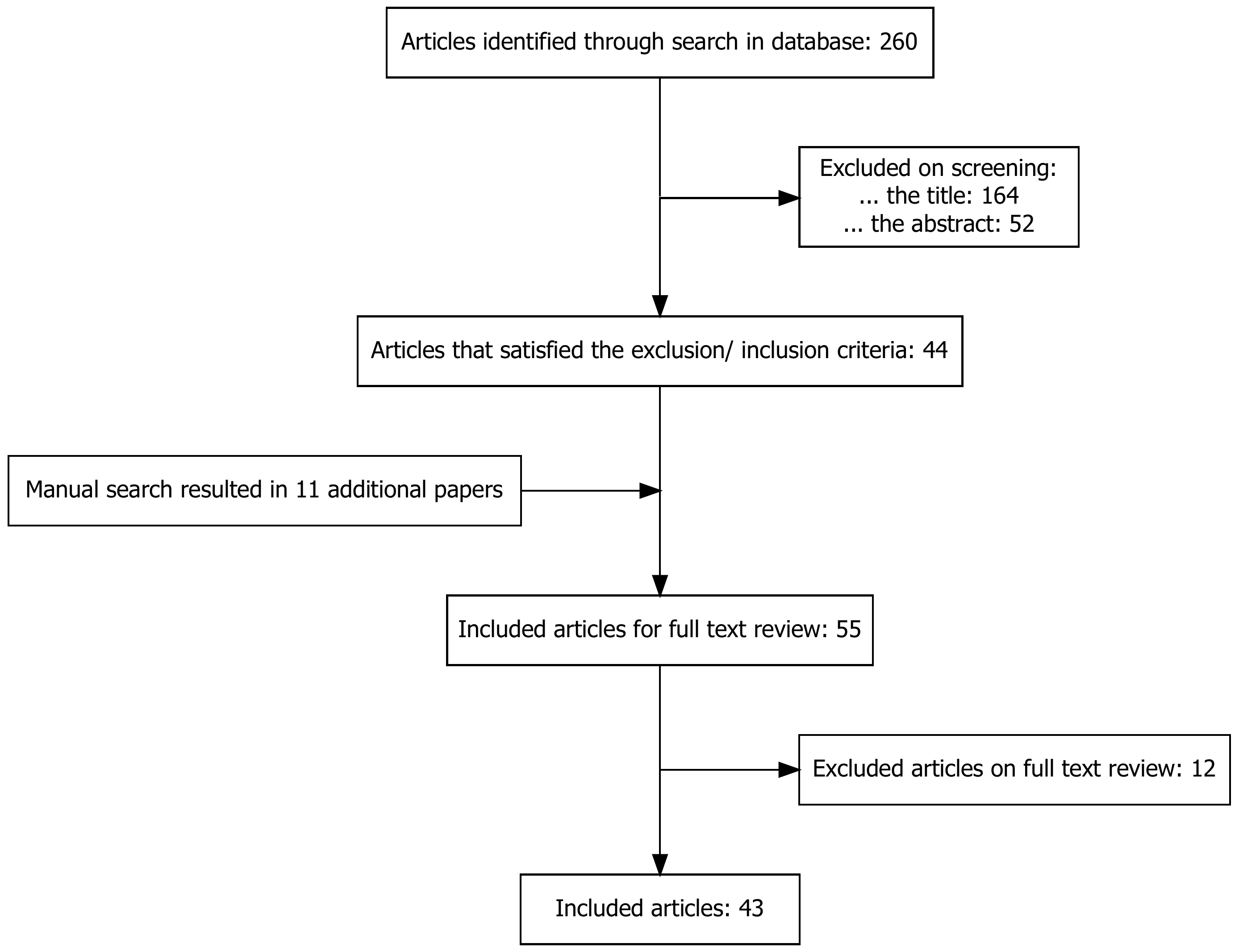}
	\caption{Flow diagram of systematic article selection process. Date of search 17/08/2021}
	\label{fig:Flowchart_ScopingReview}
\end{figure}

\subsubsection*{Data extraction}
Information from the identified articles was extracted by two independent reviewers using a standardized data extraction form (see Table 1 in the Supplementary Material 1). General information of the articles such as the year of publication, objective of paper or the type of paper (e.g. review, research article) were part of the extraction form. In addition, information on the study design (e.g. endpoints, treatment arms) was extracted. Concerning the incorporation of external/non-concurrent controls, we extracted details on the specific statistical methodology mentioned (e.g. Bayesian/Frequentist approach, covariate adjustment). Further information was collected regarding the implementation of simulation studies and the availability of code and software. Where possible, pre-specified categories were defined for each item in the extraction form. Adjudication was performed by a third reviewer in case of discrepancies. For the complete extraction form see Table 1 in the Supplementary Material 1. 

\subsubsection*{Data analysis}
The extracted information was analyzed descriptively using R Version 4.0.3. The number of articles in each pre-specified category was determined and free-text fields were summarized in listings.

\subsection*{Review of guidelines}
Our initial search was specifically on platform trials and non-concurrent controls in the strict sense. But since platform trials are still relatively new trial designs, there were so far not many guidelines available that addressed the use of non-concurrent controls. More precisely, only four platform trial specific guidelines were available: The recently published FDA guidance on "COVID-19: Master Protocols Evaluating Drugs and Biological Products for Treatment or Prevention" \cite{FDA.covid.2021}, the FDA guidance "Master Protocols: Efficient Clinical Trial Design Strategies to Expedite Development of Oncology Drugs and Biologics" \cite{FDA.master.2018} as well as the FDA guideline on "Interacting with the FDA on Complex Innovative Trial Designs for Drugs and Biological Products" \cite{FDAC.complex.2020}, and, as the only European document, “Recommendation paper on the initiation and conduct of complex clinical trials” \cite{CTFG.complex.2019}. The topic of non-concurrent controls was either not considered at all or only marginally in these guidelines. Hence, we decided to broaden the review to guidance provided on the use of external or historical controls in clinical trials and to discuss the relevance and transferability of the results to non-concurrent controls.

\subsubsection*{Search strategy}
Also this systematic guideline review was performed in accordance with the PRISMA reporting guideline's extension for scoping reviews \cite{Pagen71}. The database for our systematic review of guidelines was based on all documents available for download on 20/05/2021 from the database of the European Medicine Agency (EMA) as well as of the U.S. Food and Drug Administration (FDA). In the EMA search engine\footnote[1]{\url{https://www.ema.europa.eu/en/search/search}}, we activated the filters ``Topic = Scientific guidelines", ``Categories = Human", "Type of content = Documents" and ``Include Documents = Yes". In the FDA Guidance Documents search\footnote[2]{\url{https://www.fda.gov/regulatory-information/search-fda-guidance-document}}, we filtered the documents for ``Product = Drugs" as well as ``Product = Biologics". We then used the advanced search function of Adobe Acrobat Pro 2020 to search in all pdf documents for the terms:\\
{\small\begin{verbatim}
	non-concurrent control(s)
	concurrent control(s)
	historical control(s)
	shared control(s)
	historical borrowing
	external control(s) 
	master protocol(s)
	bayesian method(s)
	\end{verbatim}}
\ \\
Duplicates and older draft versions were excluded after the keyword search.

\subsubsection*{Identification of guideline documents}
We only included guidelines for data extraction which were guideline documents, Questions and Answers (QnAs), qualification opinion or reflection papers from EMA, ICH or FDA. We excluded documents 
1) in which external/non-concurrent controls were not discussed in the context of an inclusion into the primary analysis (e.g. the use of external/non-concurrent controls were just mentioned in the context of sample size planning);
2) in which one of the keywords and hence the use of external/non-concurrent controls was only mentioned without further recommendation or description (e.g. mentioned only in the title of a reference); 
3) in which the use of external/non-concurrent controls was only mentioned in a non-clinical or preclinical setting 
4) or in the context of (secondary) safety data analyses or meta-analyses; 
5) in which the use of external/non-concurrent controls was discussed in a medical device context.

\begin{figure}[ht]
	\centering
	\includegraphics[scale=0.3]{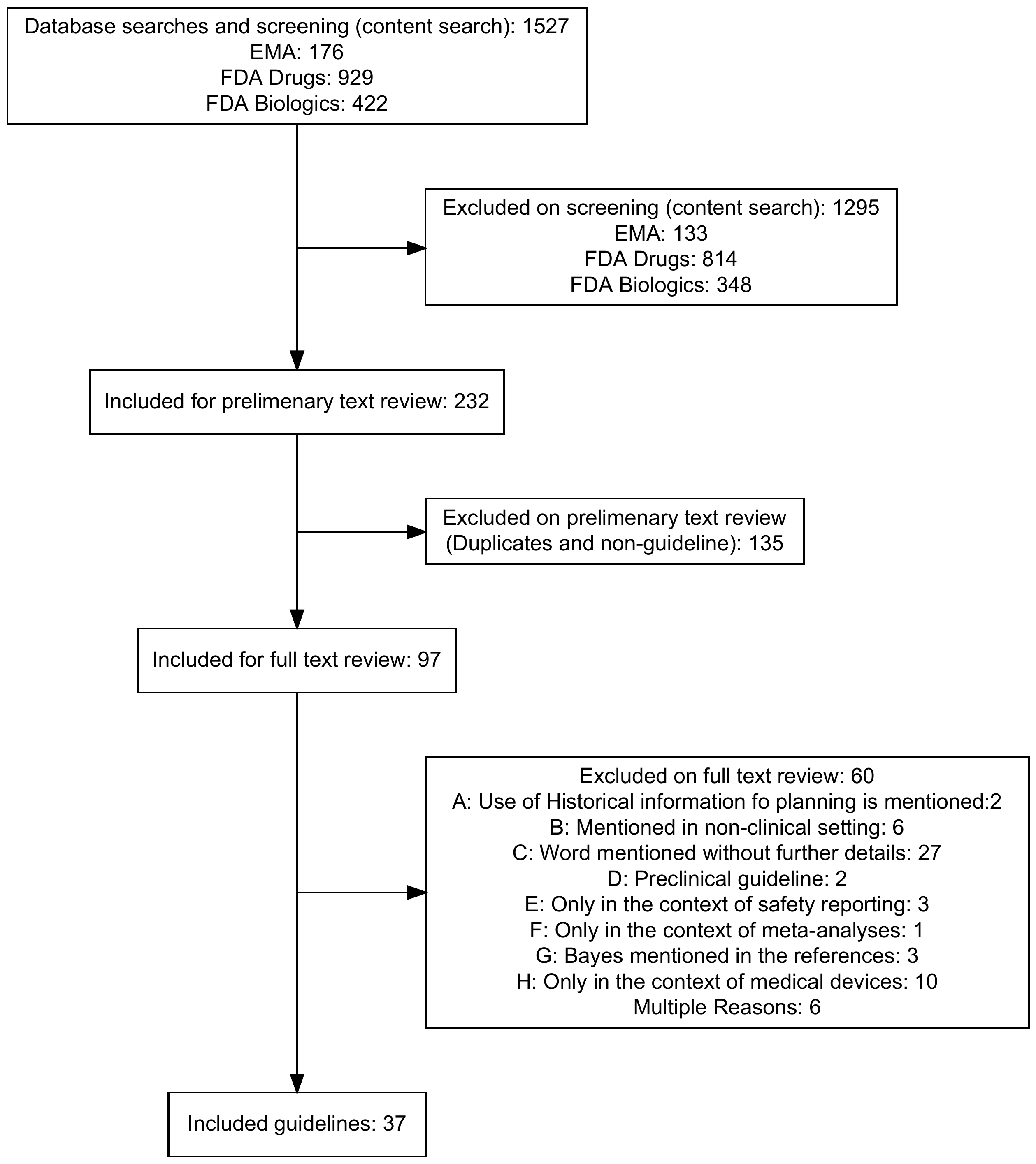}
	\caption{Flow chart of guidelines identification. Date of search 20/05/2021}
	\label{fig:Flowchart}
\end{figure}

\subsubsection*{Data extraction}
Information from the identified guidelines was extracted by two independent reviewers using a standardized data extraction form (see Table 3 in the Supplementary Material 1). General information of the guidelines such as the year of the guideline or the type of document (e.g. guideline, reflection paper) were part of the extraction form. We documented whether the guideline discussed the use of external/non-concurrent data in early or late phase and whether the guideline was focused on methodological or clinical aspects. Concerning the use of external/non-concurrent controls, we extracted details on the specific circumstances in which the use was recommended or deemed acceptable, or unacceptable, the concerns that were raised as well as the requirements for the use. Furthermore, we identified the methods mentioned for the incorporation of external/non-concurrent data, the type of inferential question addressed and whether Bayesian methods were supported. We specifically identified whether the use of non-concurrent controls, or the joint use of external and concurrent controls in platform trials was discussed in the guideline. Where possible, pre-specified items were defined for each category in the extraction form after a first pre-screening (Supplementary Material 1 Table 3). Adjudication was performed by a third reviewer in case of discrepancies.

\subsubsection*{Data analysis}
We analysed the extracted data descriptively with R Version 4.0.3. The number of guidelines in each pre-specified category was determined and free-text fields were summarized in listings.

\section*{Results of the Methods Review}

The data base search yielded 260 articles. Based on the titles 164 papers which did not address statistical methods were excluded, leaving 96 papers. We further excluded 52 articles after screening the abstracts. Hence, 44 articles satisfied the inclusion-exclusion criteria based on title and abstract screening. 
Additionally, 11 articles resulting from a manual search entered the full text review. Based on the full-text review further 12 articles were excluded leading to a total number of 43 relevant articles for which we performed the data extraction. The work-flow of the literature search is depicted in Figure \ref{fig:Flowchart_ScopingReview}. A full list of these 43 articles can be found in Table 2 in the Supplementary Material 1.

\begin{figure}[ht]
	\centering
	\includegraphics[scale=0.52]{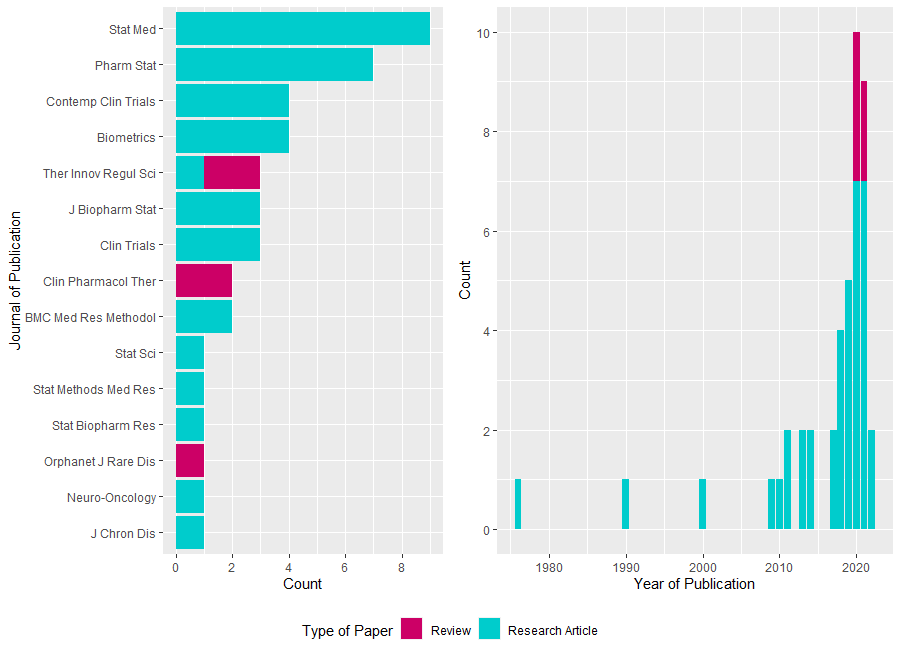}
	\caption{Journal and year of publication of the identified articles.}
	\label{fig:Plots_combined}
\end{figure}

The distribution of the year of publication of the 43 identified articles shows an increase in publications, especially in the last two years (see Figure \ref{fig:Plots_combined}). The journals with the most articles published on this topic were \textit{Statistics in Medicine} with 9/43   (21\%) articles  and \textit{Pharmaceutical Statistics} with 7/43 (16\%) articles  (see Figure \ref{fig:Plots_combined}).

Only 7/43 (16\%) of the articles focused on platform trials while 36/43 (84\%) of the articles concentrated on external controls. 28/43 (65\%) articles considered the situation of a trial with only one treatment arm. In  21/43 (49\%) of the identified articles, the considered primary endpoint was  binary, for 16/43 (37\%) of the articles a continuous endpoint was chosen, in 10/43 (23\%) it was a survival endpoint, in 2/43 (5\%) the described methodology was for count data and in 5/43 (12\%) the articles provided high-level summaries of methodologies or discussions on the use of external controls without details on specific endpoints. 
With respect to the statistical methodology, in 28/43 (65\% ) of the identified articles a Bayesian approach was used to incorporate external/non-concurrent controls while 7/43 (16\%) used a Frequentist approach and 8/43 (19\%) considered both.

We distinguish methods into two categories: \textit{downweighting-based approaches}, referring to methods that downweight the non-concurrent control data in favour of the control data either using Bayesian methods, such as meta-analytic approaches, or propensity score approaches; and  \textit{modelling-based approaches}, referring to methods that use regression models to incorporate historical or non-concurrent control data. 
The majority of the articles, 34/43 (79\%), considered a downweighting-based approach. A modelling-based approach was considered in 11/43 (26\%) of the identified papers. 
Furthermore, we found that 37/43 (86\%) of the articles discussed the potential biases and 11/43 (26\%) covered interim analyses. 31/43 (72\%) of the articles reported a simulation study or a case study and for 13/43 (30\%) articles software or code is available.

Note that some articles fall into several categories and that is why the total sums to more than 100\% in some cases. 

\subsection*{Methods to incorporate non-concurrent controls}

In this subsection, we present an overview and description of the main methods proposed in the identified articles. Most of these methods were originally proposed in the context of using historical controls and real-world evidence, but can also be applied to incorporate non-concurrent controls in platform trials.  

\subsubsection*{Test-then-pool approaches}

In the ``test-then-pool" approach, the distributions of the non-concurrent and concurrent controls are first tested for equality using a frequentist test at level $\alpha$. If the null hypothesis of equality of the distribution functions is rejected, the non-concurrent control data is discarded and a separate analysis using solely the concurrent control data  is conducted. If the hypothesis of equality can not be rejected, the non-concurrent and concurrent control data are assumed to be comparable and a pooled analysis is conducted \cite{Viele2014, Dejardin.2018}. 
In the test-then-pool approach, the significance level of the pre-test can be chosen to reflect the a priori trust in the similarity of controls. 

In the specific context of platform trials, Ren et al.  \cite{Ren2021} discussed the incorporation of non-concurrent controls in the analyses by means of a pooling approach in a trial with two treatment arms and a shared control, where one of the treatment arms enters later. The authors assumed that one treatment arm joins the platform later and compared different scenarios in which the treatment was tested against concurrent controls only, as well as against the pooled concurrent and non-concurrent controls without pre-testing the potential differences between controls. They evaluated the overall study power, defined as the probability of detecting at least one effective treatment, and the type I error control, as well as the optimal allocation ratio for the overlapping period between the two treatment arms.   Another pooling approach was introduced by Jiao et al. \cite{Jiao2019}, in which a platform trial with independent controls per treatment arm was introduced with a potential incorporation of external controls. The authors proposed a two-step test-then-pool design. In the first step, the test-then-pool approach was applied to the independent control groups in the platform trial to examine pooling of these groups to a common control. In the second step, the test-then-pool approach was applied to an external control and the platform control group (either separate or pooled control, depending on the first step).

\subsubsection*{Frequentist and Bayesian regression model approaches}

In the specific context of platform trials with continuous data and without interim analyses, Lee \& Wason \cite{Lee2020} considered linear regression models that include a factor corresponding to time to adjust for time trends when using non-concurrent controls in the final analysis. They demonstrated in a simulation study that modelling time trends by a step-wise function leads to unbiased tests, even if the true time trend is linear rather than step-wise. These models were recently further investigated by Bofill Roig et al. \cite{roig2021model} for platform trials with continuous and binary endpoints, showing that the regression model adjusting for time trends using a step-wise function relies on the assumption of equal time trends across all arms and on the correct specification of the scale of the time trends in the model. For fixed sample platform trials (without interim analysis), they show that under these assumptions the regression model gives valid treatment effect estimates and asymptotically controls the type 1 error if block randomisation is used. 

For platform trials with continuous endpoints, Saville et al. \cite{Saville2022} proposed a Bayesian generalized linear model, the so-called ``Bayesian Time-Machine''. The Time-Machine allows to model the time trends and to perform an adjusted analysis. In this approach, in order to model the potential drift over time in response, time is divided into pre-defined time ``buckets" (e.g. months or quarters) and the estimates for the time effects in different periods are smoothed using a normal dynamic linear model.  
The time-machine can achieve nearly unbiased estimates if time trends are equal in all treatments and the time buckets and priors in the model are chosen appropriately. The model can also give an improvement in terms of power and mean square error as compared to approaches that estimate the time period effect independently in each time period.

Note that a difference in these model-based approaches is the definition of the time intervals. In the frequentist models in \cite{Lee2020}, the time intervals are defined by the time at which arms enter or leave the trial, but in Saville et al., the intervals are defined based on the calendar times by means of the time ``buckets" \cite{BofillRoig2022}.

\subsubsection*{Propensity Score approaches and baseline covariates-adjustments}

Propensity score approaches have been proposed to adjust for differences between historical and concurrent controls. The scores are estimated with  regression models based on baseline covariates \cite{Schmidli2020}. 

Propensity scores were originally introduced by Rosenbaum and Rubin \cite{rosenbaum1983} in the context of causal inference in observational studies. The propensity score is the probability of being in one treatment group rather than the other given the observed baseline covariates. It is used as a balancing score, since patients with the same propensity score can be considered to have balanced covariates between the two groups. The balancing is then performed by matching, stratification, weighting or covariate adjustment using the propensity scores. 

Yuan et al. considered in \cite{yuan2019} two different approaches to augment a concurrent control arm with historical control data via propensity score matching. In \cite{chen2020} Chen et al.~proposed a propensity score-integrated composite likelihood approach for augmentation of the concurrent control arm with real-world data, in which the composite likelihood is utilized to downweight the information contributed by the external controls in each propensity score stratum. 

Covariate adjustment can also be directly implemented with regression models including relevant baseline covariates. For example, in the context of borrowing of historical control data, 
Han et al. \cite{Han2017} proposed a Bayesian hierarchical model that can incorporate patient-level baseline covariates to calibrate the exchangeability assumption between concurrent and historical control data. 

Collignon et al. \cite{Collignon2020} discussed how the inter-study variation might be useful to quantify the amount of information that historical controls can provide, and opted for a clustered allocation design as the closest to a randomised trial.
%

\subsubsection*{Power prior and Commensurate Power prior}
The Bayesian power prior approaches  discount the historical data by a power parameter to account for potential differences between historical and concurrent control data. In their seminal paper, Ibrahim and Chen \cite{Ibrahim2000} proposed power priors in the context of regression models. For the prior specification of the regression coefficients, available historical data is used together with a scalar weight parameter that quantifies the uncertainty in the historical control data. The choice of such a parameter determines the weight of the historical data to be incorporated into the current study and hence, has implications on the operating characteristics of the trial. Although the choice of the value of this parameter is thus essential, it is often challenging to specify.  

Duan et al. \cite{duan2006} and Neuenschwander et al. \cite{Neuenschwander2009} proposed a modified power prior approach in which the weight parameter is considered an unknown parameter. This modified prior aligns with the Bayesian rationale and thus, introduces a prior for this weight. Banbeta et al. extended in \cite{Banbeta2019} the modified power prior to incorporating multiple historical control arms. 
Gravestock et al. proposed in \cite{gravestock2017} an empirical Bayes approach to estimate the weight parameter based on the observed data. Bennett et al. \cite{Bennett2021} proposed a different approach to derive the weight parameter. They focused on adaptive designs with binary endpoints in which historical controls can replace the concurrent controls when there is an agreement between the historical and concurrent control data at the interim analysis. The authors \citep{Bennett2021} proposed two Bayesian methods for assessing the agreement between historical and concurrent control: the first method is based on an equivalence probability weight and the second on a weight based on tail area probabilities. 

The commensurate power prior \cite{Bennett2021,Hobbs2011} is another adaptive modification of the original power prior formulation \cite{Ibrahim2000} that uses conditional prior distributions for the concurrent controls, which adjusts the weight parameter through a measure of commensurability.  In the specific context of platform trials, Normington et al. \cite{Normington.2020} also considered a trial with independent controls per treatment arm. Instead of borrowing data from an external control, the authors proposed a commensurate prior to borrow control data from the independent controls on the same platform. In comparison to an all-or-nothing approach (pool or discard completely), their design did not perform worse in simulations in terms of treatment effect bias (if control data was wrongly included) or longer study duration (if control data was wrongly discarded).

\subsubsection*{Hierarchical models}
Meta-analytic-predictive (MAP) prior approaches account  for heterogeneity by assuming exchangeability among the historical and concurrent control parameters and explicitly model the between-trial variation. This method performs  a prediction of the control effect in a target clinical trial from historical control data using random-effects meta-analytic methods \cite{Schmidli2020}. Schmidli et al. \cite{Schmidli2014} defined a robust extension of the MAP (R-MAP) prior to allow for further discounting of historical data in the case of extreme discordance between the historical and concurrent control data. This prior is a mixture prior defined by two components, a MAP prior based on the historical data and a weakly-informative prior. Additionally, they proposed an adaptive design where in the interim analysis the agreement between historical controls and concurrent controls is evaluated. If the data is in agreement, then fewer concurrent controls will be recruited in the second stage. 

More recently, Hupf et al. extended in \cite{Hupf2021} the MAP prior approach and proposed the Bayesian semiparametric MAP prior (BaSe-MAP). In the BaSe-MAP approach the random effects in the MAP prior are modelled nonparametrically as a Dirichlet process mixture of Gaussian distributions centered on a common mean. The BaSe-MAP borrows more conservatively than the other priors, but its performance in terms of frequentist operating characteristics is similar or better than the MAP and robust MAP methods.

Wang \cite{Wang2022} proposed a new Bayesian model for adjusting for time trends in platform trials. 
The approach is based on pooling control data, using dynamic borrowing depending on how similar concurrent and non-concurrent controls are. In their setting,  the standard of care control arm was permitted to be changed over time in the platform. Therefore, comparisons to the control arm were subject to change and therefore, changes over different recruitment times were expected. The authors proposed an extension of the R-MAP approach to accommodate these changes. Specifically, the approach uses mixture priors that link the most recent non-concurrent controls with those furthest away from the concurrent, and then using this to decide how far in time the non-concurrent controls should be pooled. The methods are illustrated by mean of examples based on e.g. trials of Ebola virus disease therapeutics.

\subsubsection*{Elastic prior}
The elastic prior method \cite{Jiang2021} uses a so-called elastic function, which is basically a congruence measure mapped to (0,1). This elastic function measures the strength of evidence for the congruence between concurrent and historical data (e.g. a test statistic). The elastic function is constructed to satisfy a set of pre-specified criteria such that the resulting prior will strongly borrow information when historical and trial data are not in conflict with each other, but refrain from information borrowing when historical and trial data are in-congruent by inflating the variance of the prior distribution. The method was extended in \cite{Zhang2021} to biosimilar studies. 

\subsubsection*{Pocock's random bias model}
In his seminal paper \cite{Pocock1976}, Pocock proposed a Bayesian statistical method that accounts for the difference in the historical and concurrent controls by means of a bias term, which is assumed to be a normally distributed random variable with mean zero. The method is equivalent to the commensurate prior in \cite{Hobbs2011} in case of a single historical trial, except that the between-study variance is not estimated but set in advance. In addition, the random bias model is similar to the MAP and R-MAP approach. For further discussions on the similarities of these methods see \cite{rosmalen2018} and \cite{callegaro2021}.

\subsubsection*{Discussion and comparison of methods}

Jiao et al. \cite{Jiao2019} compared several strategies to borrow historical external control data in the context of platform trials, including test-then-pool, dynamic pooling and MAP-prior, through a simulation study with respect to the type I error of rejecting the null hypothesis of a new added arm. They showed that when using the pooled approach, the type I error (T1E) might be strongly inflated if there are positive time trends. In the test-then-pool approach, there may be a T1E inflation but it depends on the significance level used in the pretest. In the same line, the dynamic pooling and the MAP approach might lead to an inflation when there are positive time trends depending on the value of the weight parameters used. A further note is that test-then-pool and MAP approaches have  bounded inflation of the T1E, while for others, the T1E goes to 1 when the differences between the concurrent and non-concurrent controls become larger. 
Isogawa et al. compared in \cite{Isogawa2020} MAP prior and the Power Prior approaches for the incorporation of historical control data in clinical trials with a binary endpoint. They summarize the results of their simulation study with the conclusion that if importance is attached to control T1E, the MAP approach based on a normal-normal hierarchical model may be preferred, while the power prior borrows in general more and hence, has larger power at the cost of a larger T1E inflation in case of conflict of historical and concurrent control data.

In Burger et al. \cite{Burger2021}, the authors elaborated on different sources of bias when using external controls including but not limited to calendar time bias, selection bias and regional bias. Platform trials were mentioned as one of the potential applications of external controls. They acknowledged that non-concurrent controls in platform trials can not be entirely considered ``external" but are subject to similar biases as external controls, especially to time trends in the control group. As mentioned by Lee \& Wason \cite{Lee2020}, or Saville et al. \cite{Saville2022}, regression models could be used to accommodate the time trends and lead to unbiased estimators and T1E control under the assumption of equal time trends on the model scale \cite{roig2021model}. 
In \cite{Jahanshahi2021}, Jahanshahi et al.~aim to better understand the use of external controls to support product development and approval. They reviewed FDA regulatory approval decisions between 2000 and 2019 for drug and biologic products to identify pivotal studies that leveraged external controls, with a focus on selected therapeutic areas. Although the scope of the paper is not to review the statistical methods on the use of external controls, they highlighted some of the methods and approaches often used. They referred to three statistical approaches often used to adjust for baseline imbalances: matching, covariate adjustment, and stratification. They discussed advantages and disadvantages of using propensity scores to match or stratify based on the score. As an alternative, they mentioned analysis approaches to evaluate the consistency among results using different sources of external control data. 

\section*{Results of the guideline review}

Overall, we found 1527 documents from the EMA and FDA database (see Figure \ref{fig:Flowchart}): 176 documents from the EMA database and 1351 documents from the FDA website. In all these documents, we searched for the above defined keywords which resulted in 232 documents. In a first filtering step, we excluded duplicates and old drafts for which a newer draft or final version was already included in the downloaded documents. As a result, 97 documents were included for a full text review. 
Based on the inclusion and exclusion criteria, 60 guidelines were excluded and 37 guidelines were identified as relevant for our final review and data extraction (for a full list of included guidelines see Supplementary Material 1 Table 4).
As already mentioned above, only four platform trial specific guidelines were available at the time of extraction\citep{FDA.covid.2021,FDA.master.2018,FDAC.complex.2020,CTFG.complex.2019}. All four documents raised the issue of potential time drifts. For this reason, the FDA guideline for Master Protocols in COVID-19 takes a clear position against the use of non-concurrent controls in this context. The other documents did not provide further
guidance or details. 
Therefore, as mentioned above, we broadened our search to guidance provided on the use of external or historical controls in clinical trials and discuss the relevance and transferability of the results to non-concurrent controls in platform trials. The discussion of the regulatory guidelines is the interpretation of the authors only and cannot always be directly derived from the guidance documents.

\subsubsection*{Circumstances in which the use of external controls is potentially acceptable}
\begin{figure}[ht]
	\centering
	\includegraphics[scale=0.32]{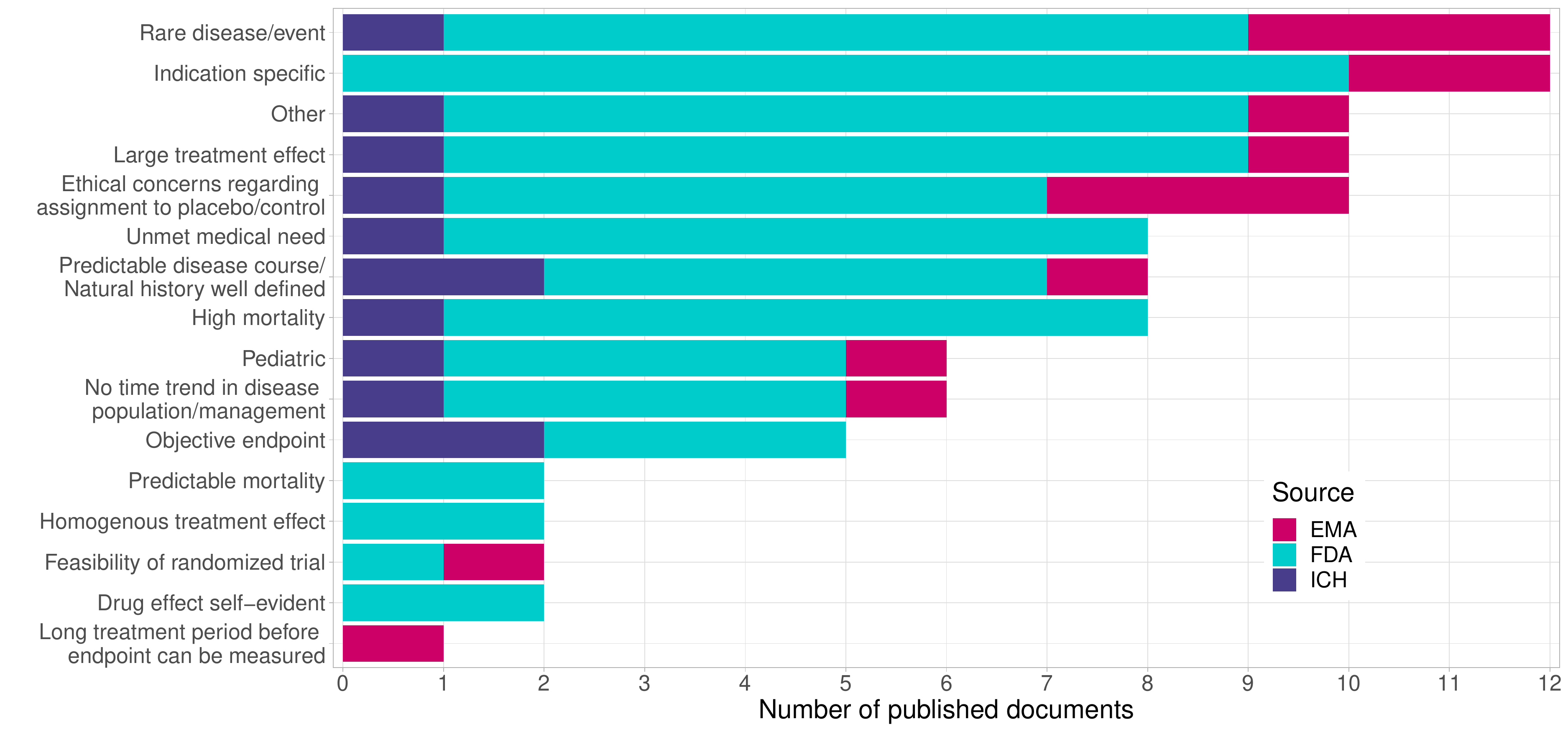}
	\caption{Circumstances in which the use of external/historical/non-concurrent control is recommended or deemed acceptable. Categories are not mutually exclusive. One guideline can be present in multiple circumstance categories. Other was selected in case a guideline mentioned a circumstance which was not pre-specified such as a reference to ICH E10.}
	\label{fig:Q7_Circumstances}
\end{figure}

Based on a first text screening of the guidelines, we pre-specified the following circumstances in which the use of external controls might be acceptable: the targeted indication is related to a rare disease, the trial is in a pediatric context, there is an unmet medical need, the indication is related to a high mortality, a long treatment period is needed before the endpoint can be measured, a large treatment effect to be expected, no time trend in disease population or management is expected, a homogeneous treatment effect to be expected as well as there exist ethical concerns regarding assignment to the control group (see Figure \ref{fig:Q7_Circumstances}). The circumstance that was mentioned the most in the  identified guideline documents was “rare disease” or was an "indication specific" concern. But also a “large treatment effect” or “ethical concerns” were mentioned quite often. All of these results relate quite generally to historical or external controls. If we look specifically at these circumstances under the focus of non-concurrent controls, we see that not all of these circumstances are one to one transferable to the question "when to use or not use non-concurrent controls" for the final analysis in a platform trial. But “large treatment effect”, “no time trend in disease course” as well as an “objective endpoint” could, for example, be circumstances in which also the use of non-concurrent controls in a platform trial might be me more appropriate. 

\subsubsection*{Methods mentioned for the use}
\begin{figure}[ht]
	\centering
	\includegraphics[scale=0.32]{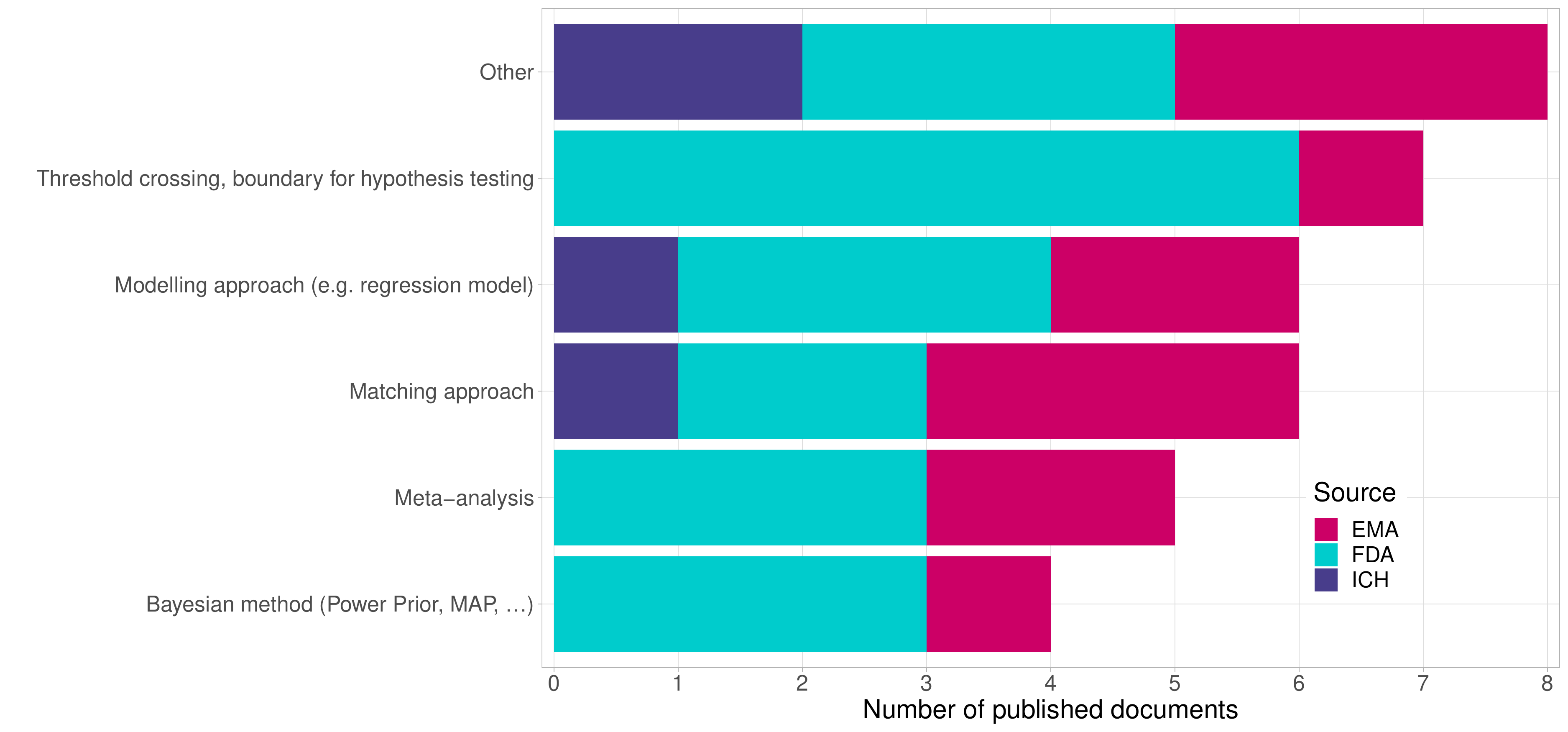}
	\caption{Methods mentioned for the use of historical/external/non-concurrent controls. Categories are not mutually exclusive. One guideline can be present in multiple methods categories. Other was selected in case a guideline mentioned a potential use of external data without a specific methodology (6 guidelines), or a methodology which was not pre-specified such as extrapolation or simulation (2 guidelines).}
	\label{fig:Q9_Methods}
\end{figure}

Within the methods one can of course distinguish between Bayesian as well as Frequentist methods. 
We furthermore decided to differentiate between (Frequentist) regression model approaches, matching approaches, a classical Meta-Analysis approach as well as classical threshold crossing whereby the threshold is based on historical control data (see Figure \ref{fig:Q9_Methods}). As an overall conclusion we can say, that the guidelines remain vague rather than instructive on the methods. None of the methods was mentioned conspicuously more often than another; and none of the methods was described or discussed in detail. More or less, if methods were mentioned at all, they were mentioned only in a passing sentence. Regarding the relevance of non-concurrent controls: all methods applicable are also applicable to non-concurrent controls. Due to overlapping time-periods, modelling approaches which attempt to model a time trend fit here probably even much better.  

\subsubsection*{Concerns} 
\begin{figure}[ht]
	\centering
	\includegraphics[scale=0.32]{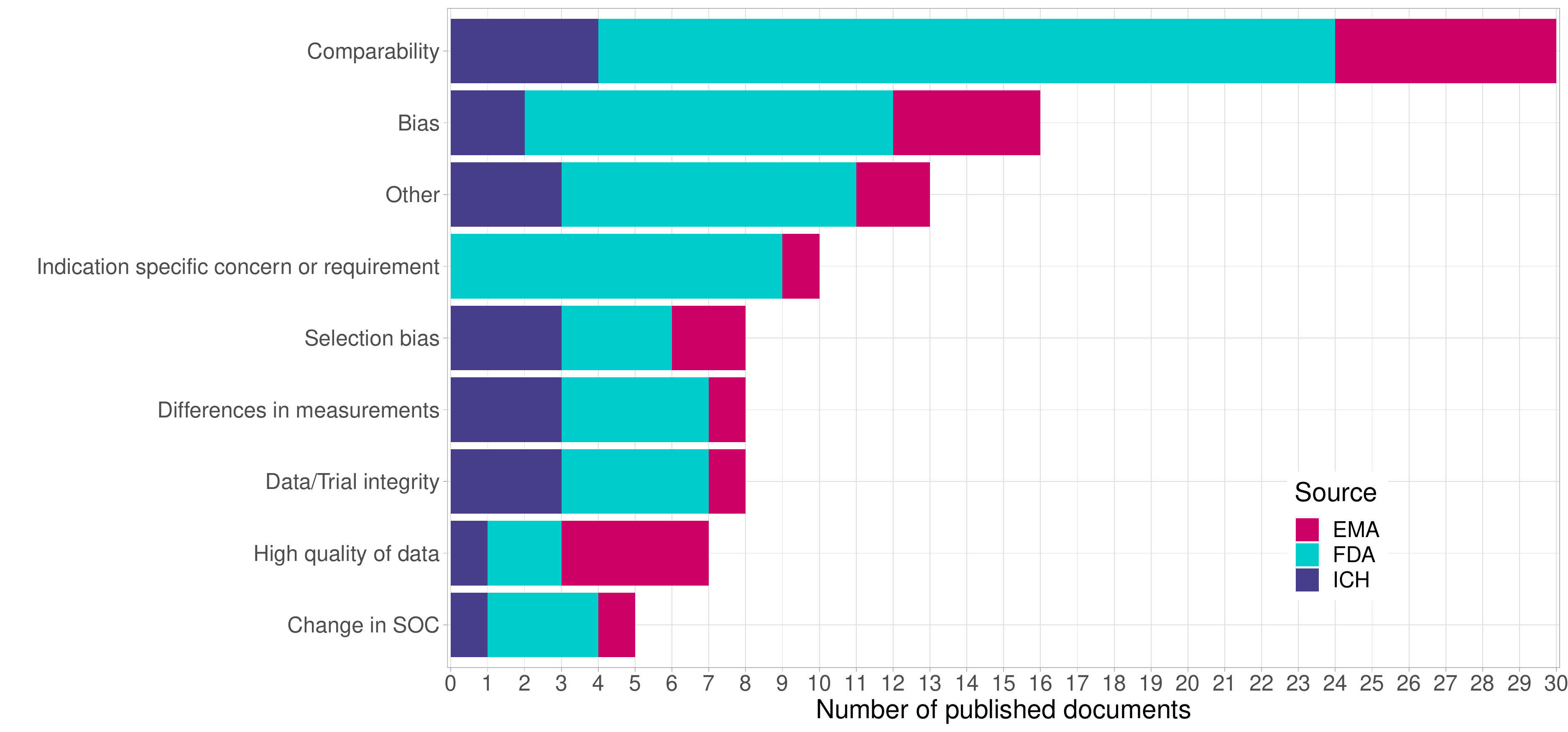}
	\caption{Concerns raised with the use of historical/external/non-concurrent controls. Categories are not mutually exclusive. One guideline can be present in multiple concern categories. Other was selected in case a guideline mentioned a concern which was not pre-specified such as missing unknown important prognostic factors in historical controls.}
	\label{fig:Q8_Concerns_Requirement}
\end{figure}

Non-comparability and bias were raised most often as the general concerns with external controls. Selection bias, differences in measurements, data/trial integrity and changes in standard of care were raised several times as more specific concerns that can lead to non-comparability and bias (see Figure \ref{fig:Q8_Concerns_Requirement}). For a detailed explanation of these terms, refer to \cite{Burger2021}. Evaluating the risks for these issues for non-concurrent controls in platform trials, the authors see the risks of selection bias and differences in measurement as significantly reduced: No selection of specific data sources and data points is done so that there cannot be any willful selection bias, nor is publication bias an issue. In platform trials, the definition of measurement systems and their quality assurance is part of the clinical trial planning, such that the use of the same instruments and standard operation procedures throughout the trial can be mandated, and e.g. for newly developed measurements, shift and drift and operator bias can be taken care of. For the risks on data and trial integrity in platform trials, the authors see platform trials comparable to adaptive trials rather than to trials using external controls, that means, they exist but can be mitigated more easily in platform trials than in trials with external controls: Changes in the behaviors of investigators, patients and carers upon revelation of results can be mitigated in platform trials by good planning and precaution measures in masking and information dissemination. This is a big advantage over external controls, where there is very limited masking possible, and often knowledge on the external controls is available to many stakeholders of the trial. Changes in the standard of care can occur in the course of a platform trial, and in that case, this clearly limits the comparability of non-concurrent controls just as it would be the case for any external control group. The advantage over external controls is the level of transparency on the timing and extent of such a change in platform trials. This allows informed decision making on which non-concurrent controls to use or not to use.

\subsubsection*{Requirements} 
Requirements mentioned in guidelines mirror the concerns that are raised: the populations have to be comparable, and the data of high quality (see Figure \ref{fig:Q8_Concerns_Requirement}).

\subsubsection*{Indication specific recommendations} 
The most instructive guidance can be found on the use of external controls for specific indications. There is discouragement if the disease is very heterogeneous (Amyotrophic Lateral Sclerosis \cite{FDACDER.Amyotrophic.2019}, Acute Myeloid Leukemia \cite{FDA.Acute.Myeloma.2020}), when the disease is known to change over time (Influenza \cite{FDACDER.Influenza.2011}, COVID-19 \cite{FDA.covid.2021}), when bias or changes over time are suspected for procedures in general and endpoint evaluation (functional tests in Duchenne Muscular Dystrophy \cite{FDA.duchenne.2018}, evaluation techniques for objective response rate and progression free survival in oncology \cite{FDA.endpoints.2018}, dietary management \cite{FDA.dietary.2018}), and where standard of care is changing rapidly (COVID-19 \cite{FDA.covid.2021}, Knee Cartilage \cite{FDA.IDE.2010}). On the positive side, Table \ref{tab:Indication_recommendation} gives examples of recommendations on including external controls. In case of the intravenous immunoglobulin replacement therapy trials, the FDA switched from a clear randomized clinical trial recommendation to a historically controlled recommendation between 1999 and 2000, and the process is described in more detail in \cite{Schroeder.2012}. These examples show how the details of trial designs help in deciding about the variance-bias trade-off for external controls, and raise the hope that instructive guidance can be developed for platform trials for specific scientific questions.

\begin{table}[h!] 
	\caption{Recommendations for the use of external controls}
	\label{tab:Indication_recommendation}
	\centering
	\begin{tabular}{ll}
		\hline
		\textbf{Document title} & \textbf{Citation} \\
		\hline
		Chronic Hepatitis C Virus Infection: &  The  primary  efficacy  comparison  (either superiority or \\
		Developing Direct-Acting Antiviral  &  noninferiority  depending  on regimen studied) should \\
		Drugs for Treatment &   be  to  a historical  reference of  a recommended  HCV  \\
		& treatment  regimen  rather than  a  comparison to  those  \\
		& receiving placebo (in the  deferred  treatment  arm)   \\
		& because it is expected that no patient will respond   \\
		& virologically while receiving  placebo. \\
		\hline
		Antibacterial Therapies for Patients  &  we recommend randomizing at least a small number  \\
		With an Unmet Medical Need for the & of patients to the active control (e.g., through \\
		Treatment of Serious Bacterial  & disproportionate randomization of 4:1), if feasible and \\
		Diseases & ethical based on an active control considered to be  \\
		& best-available therapy. This will allow for an assessment      \\
		& of the comparability of the external control to the trial    \\
		& population. Frequentist and Bayesian statistical    \\
		& methods can then be  used to combine external control   \\
		& data with data from the patients randomized to the    \\
		& active control in assessing differences between   \\
		& treatment groups for the primary comparison.\\
		\hline
		Reflection paper on the regulatory  & If  the  epidemiology  of  SARS-CoV-2  indicates  that  it is  \\
		requirements for vaccines intended  &  no longer  in  the  best interest of subjects  to receive    \\ 
		to provide protection against  &  primary vaccination  with the  parent  vaccine,  an      \\
		variant strain(s) of SARS-CoV-2 &  alternative approach to  immunobridge  from the  efficacy     \\
		& previously  documented with the  parent  vaccine  to  the     \\
		& variant  vaccine could be  a comparison between  immune    \\
		& responses  elicited  by  primary vaccination  with  the         \\
		& variant vaccine against the  variant  strain and prior  data     \\
		& on  the immune  response  elicited by   primary vaccination      \\
		& with the parent  vaccine against  the  parent strain.  \\
		\hline
		Safety, Efficacy, and Pharmacokinetic & The protocol should prospectively define the study \\
		Studies to Support Marketing of as & analyses. We expect that the data analyses presented \\
		Immune Globulin Intravenous (Human)  & in the BLA will be consistent with the analytical plan \\
		Replacement Therapy for Primary &   submitted to the IND. Based on our examination of  \\
		Humoral Immunodeficiency &historical data, we believe that a statistical  \\
		& demonstration of a serious infection rate per person-\\
		& year less than 1.0 is adequate to provide substantial    \\
		& evidence of efficacy. You  may test the null hypothesis  \\
		& that the serious infection  rate  is greater than or   \\
		& equal to 1.0 per person-year at the 0.01 level of   \\
		& significance or, equivalently, the upper  one-sided 99   \\
		& Percent confidence limit would be less than 1.0\\
		\hline
	\end{tabular}
\end{table}


\section*{Conclusions} 

We conducted a systematic search for statistical methods to incorporate non-concurrent controls when analysing platform trials. 
We identified methods originally proposed in the context of the use of historical and real-world data in clinical trials, as well as methods proposed for platform trials utilising non-concurrent controls. The approaches can be classified in two broad categories: 
\textit{downweighting-based approaches}, which concerns methods that downweight the non-concurrent control data in favour of the concurrent control data depending on how similar the concurrent and non-concurrent control data are; and 
\textit{modelling-based approaches}, which are methods that use regression models to incorporate historical or non-concurrent control data while adjusting for time trends by incorporating time as covariate in the model. 
Downweighting-based approaches have been widely discussed in the context of historical data. The validity of the results when using these downweighting approaches might strongly depend on the pre-specified parameters and assumptions made. Besides, none of these approaches control the type 1 error in all the scenarios. For example, when using down-weighting approaches with Bayesian methods, Kopp-Schneider et al. \cite{kopp2020power} have shown that  strict control of the type 1 error rate is not possible when incorporating external information. In the case of model-based approaches, unbiasedness and type 1 error control depend on the assumption of equal time trends in all arms on the model scale. Some methods also use  other covariates to adjust for temporal drifts. 
When using covariate-adjustment approaches, the unbiasedness and error control will depend on whether all covariates causing time trends are included in the model and collected in the data. 

Recently,  the estimand framework \cite{ICH_9R1} has become an important part of clinical trial protocols. Of further note is the fact that the estimand, defined as the target of estimation, is derived from the trial objective and is therefore not directly affected by the use of non-concurrent controls. 
However, the inclusion of controls other than concurrent controls is an important aspect when discussing if the estimators are aligned to the estimand. Also, the methods considered to incorporate them, as 
the properties of  estimators will depend on how non-concurrent controls are employed in the estimation of treatment effects \cite{collignon2022estimands}. 

With regard to the guidelines, it can be stated that current guidelines are rather vague when it comes to recommendations and considerations on using non-concurrent controls in a platform trial. Only four guidelines specifically for platform trials were available at the time of our search. Therefore, we broadened our search to not only non-concurrent controls in platform trials but to the general use of external/historical data in clinical trials. We discussed the relevance and transferability of our results to non-concurrent controls. The listed requirements and concerns regarding the use of external controls in the guidelines under consideration are related but cannot be applied directly on the use of non-concurrent controls. Despite the broadening, there was still a lack of clear guidance on which statistical methods would be more or less appropriate to incorporate external data for regulatory decision making.  Issues about the ``quality of historical data" do usually not arise with non-concurrent controls in platform trials. Furthermore, the concerns about ``selection bias" and ``changes in measurement" are reduced. The concern about ``comparability" is decreased when inclusion and exclusion criteria are identical for concurrent and non-concurrent controls. However, that alone does not guarantee ``comparability", for example, new treatments could be added to the platform or intermediate results may become publicly available such that different patients are attracted than before. Furthermore, the concerns about ``change in standard of care" as well as ``time trends" of unknown origin are still an issue. Due to this, the FDA guideline for Master protocols in COVID-19 takes a clear position against the use of non-concurrent controls in this context. At the time of writing this article, the EMA \citep{EMA.QnA.2022} has published a QnA on complex clinical trials in May 2022. The issue of time trends is also raised in the document and a reference to ICH E10 \cite{guideline2000choice} is made for the choice and justification of a control group. However, it is also acknowledged that platform trials are typically more complex than what is mentioned in ICH E10. Additional sources of bias may arise due to the complex features of the trial such as the introduction of new treatment arms in the trial at different time points or the use of shared controls. Especially the use of non-concurrent controls may affect trial interpretability. Therefore, early interaction with regulators \cite{regnstrom2010factors} is also recommended in the QnA  \citep{EMA.QnA.2022}. Similarly to  discussions a decade ago on which adaptations \cite{elsasser2014adaptive, collignon2018adaptive, bauer2016twenty} are useful or not, it is expected that regulatory experience and acceptance of NCC methods as supportive or primary analysis will probably grow over the next years. This scoping review was based on publicly available documents, but companies may already have had more elaborated discussions with regulators, e.g. via EMA scientific advice and protocol assistance procedures. Hence, a review on recent scientific advices would be of interest for future research as it usually takes some time until arising issues are reflected in regulatory guidance documents. The EMA has just released a new concept paper on platform \citep{EMA.ConcPlat.2022} trials announcing that in the upcoming years a reflection paper also addressing the issue of non-concurrent control data should be addressed complementing the already existing guidelines.

One other key finding of our systematic guideline search is that indication-specific guidelines may have the highest potential to be instructive, since the question “To use or not to use” strongly depends on the specific indication and trial setting. For example, for rare diseases and indications less prone to changes over time, there might be a higher willingness to utilize NCC data within a platform trial compared to ``broad" and dynamic indications.

\newpage

	\section*{Acknowledgements}

	The authors also wish to thank Peter Mesenbrink and Ekkehard Glimm for useful discussions, and comments on the draft manuscript. Furthermore, we would like to thank Natalia Muhlemann for the help in discussions and review of guidelines.

	\section*{Funding}
	
	EU-PEARL (EU Patient-cEntric clinicAl tRial pLatforms) project has received funding from the Innovative Medicines Initiative (IMI) 2 Joint Undertaking (JU) under grant agreement No 853966. This Joint Undertaking receives support from the European Union’s Horizon 2020 research and innovation programme and EFPIA and Children’s Tumor Foundation, Global Alliance for TB Drug Development non-profit organisation, Springworks Therapeutics Inc. This publication reflects the authors’ views. Neither IMI nor the European Union, EFPIA, or any Associated Partners are responsible for any use that may be made of the information contained herein.  
	
	\section*{Competing interests/Disclaimer}
	The authors declare that they have no competing interests regarding the content of this article.
	The views expressed in this article are the personal views of the authors and may not be understood or quoted as being made on behalf
	of or reflecting the position of the Paul-Ehrlich-Institut, Cytel or any other of the institutions involved. 
	Also, note that the discussion of regulatory guidelines is the interpretation of the authors only and cannot be directly derived from the guidance documents. 
	
	\section*{Authors' contributions}
	MBR and MP conceived and designed the research, MBR and KH jointly led the work, designed the review, and drafted the article. MBR, KH, UG, CB and QN participated in the screening process. 
	CB and QN analysed the data regarding the methods and guidelines reviews, respectively.  
	All authors contributed to the interpretation of findings, critically reviewed and edited the manuscript. All authors critically appraised the final manuscript.
	
	
	\section*{Additional Files}
	In the supplementary material, the process for the methods review, as well as the guideline review are described in detail. The list of pre-defined articles and the forms for the information extraction process are included. Besides, a list of the included articles and guidelines for extraction are presented. 


\bibliographystyle{biorefs}
\bibliography{bmc_article} 
\newpage


\newpage



\section*{Supplementary material}

# Supplementary material of "On the use of non-concurrent controls in platform trials: A scoping review"

Marta Bofill Roig [*,1], Cora Burgwinkel[1,2], Ursula Garczarek[3], Franz Koenig[1], Martin Posch[1], Quynh Nguyen[2], and Katharina Hees[2]

[1]*Section for Medical Statistics, Center for Medical Statistics, Informatics, and Intelligent Systems, Medical University of Vienna, Vienna*
[2]*Paul-Ehrlich Institut, Department of Biostatistics, Langen*
[3]*Cytel Inc., Strategic Consulting, Hagen*

In this supplementary material, the process for the methods review, as well as the guideline review are described in detail. The list of pre-defined articles and the forms for the information extraction process are also included.

# 1 Additional information regarding the methods review

In this section, we describe the protocol followed for the methods review, and present the extraction form used for the information extraction in the systematic review and list of included publications.

## 1.1 Protocol Methods review

A systematic search was carried out in the PubMed database and the identified articles were supplemented with manually searched papers. The following keywords were considered important and were used for the query: non-concurrent control(s), concurrent control(s), historical control(s), shared control(s), historical borrowing, external control(s). To identify the relevant methods that incorporate non-concurrent control data in PubMed, the following query was performed:

```
("non concurrent control*"[Title/Abstract] OR "control arm*"[Title]
```

*marta.bofillroig@meduniwien.ac.at

```
  OR "concurrent control*"[Title] OR "historical control*"[Title]
  OR "external control*"[Title] OR "shared control*"[Title/Abstract])
AND
  ("design*"[Title/Abstract] OR "stud*"[Title/Abstract]
  OR "platform trial*"[Title/Abstract]
  OR "master protocol*"[Title/Abstract])
AND
  ("trial*"[Title/Abstract] OR "clinical"[Title/Abstract])
```

Several searches which included, for example, the term "historical control*" in the abstract, resulted in such a large number of articles that it was infeasible to review. As result, a narrower search was performed (i.e. search term only included in the title) which prioritized the relevance of the articles in terms of methodology. Besides, it was of importance that the following a priori defined list of articles of interest was found by our search and that the most relevant articles reviewing historical control methods were also included:

- Including non-concurrent control patients in the analysis of platform trials: is it worth it? Lee and Wason (2020) [1]

- Beyond Randomized Clinical Trials: Use of External Controls, Schmidli et al. (2020) [2]

- Utilizing shared internal control arms and historical information in small-sized platform clinical trials, Jiao et al. (2019) [3]

- Robust meta-analytic-predictive priors in clinical trials with historical control information, Schmidli et al. (2014) [4]

- Use of historical control data for assessing treatment effects in clinical trials, Viele et al. (2014) [5]

- Incorporating historical control data in planning phase II clinical trials, Thall and Simon (1990) [6]

This study is reported in accordance with the PRISMA reporting guideline's extension for scoping reviews [7]. The query led to 260 articles (date of search 17/08/2021) and the articles were included if one of the following inclusion criteria was fulfilled:
1) The description of a method proposed to include external/non-concurrent controls and concurrent controls in clinical trials.
2) The article considered an application of a method which included external/ non-concurrent controls together with concurrent controls in a clinical trial context with a detailed description of the method used.
3) The article is an overview of several methods to include external controls (e.g. review article).
4) The article is about the comparison of several methods (e.g. via simulation studies) to include external controls.

The article was not considered if:
1) The article is a duplicate.
2) The article focuses on shared controls but not on the inclusion of external controls.
3) The term external control is used in another context.
4) The article focuses on a clinical trial using external controls, but not on the methods.

The screening of the articles was performed by two reviewers. In the first step, article titles were screened by a reviewer and selected based on the inclusion and exclusion criteria above. Then, the article abstracts were screened and selected by a reviewer based on the same inclusion and exclusion criteria. In the next step, the selected articles were fully read and screened by two reviewers based on the inclusion criteria and exclusion criteria.

Regarding the information extraction process, two independent reviewers read the identified articles and extracted the information using a standardized data extraction form (see Table 1). Where possible, pre-specified categories were defined for each item in the extraction form. As one can see in Table 1, information about the publication such as year and journal of publication, objective of article and type of article (e.g. review, discussion, research article) were extracted. Besides, information concerning the study design, such as type of control and type of endpoints were identified. Regarding the incorporation of external/internal controls, details on the statistical methodology, such as covariate adjustment and interim analyses, were collected. Further information was gathered concerning the implementation of simulation and/or case studies and the availability of code and software. After the information extraction, a third person, the adjudicator, compared the extracted information and checked for discrepancies. In the end, the number of articles in each pre-specified category was determined and free-text fields were summarized in listings.

Following the selection steps above, 44 articles were included for the full text review. Further 11 articles were manually added (see list above), such that in total 55 articles were included in the full text review. During the full text review, 12 articles were excluded resulting in 43 articles for the information extraction process (for a full list of included articles see Table 2).

Table 1: Extraction form for systematic review

| Question | Possible options or [Data type] |
| --- | --- |
| **Publication** | |
| 1. Title of article | [Text] |
| 2. First author | [Text] |
| 3. Year of Publication | [Integer] |
| 4. DOI | [Integer] |
| 5. Journal of publication | [Text] |
| 6. Objective of article | [Text] |
| 7. Type of article | A/ B/ C [A: Review; B: Discussion; C: Research Article] |
| **Study design** | |
| 8. Is the article focused on clinical trials? If yes | Yes/ No |
| 8a. Does it focus on platform studies? | Yes/ No |
| 9. Population in the concurrent/ actual trial | A/ B/ C [A: Several populations are included; B: Only one population is included; C: Not specified] |
| 10. Type of control | A/ B [A: External controls; B: Internal controls] |
| 11. Number of treatment arms (control arm not included) | A/ B/ C/ D [A: 1 arm; B: 2 arms; C: +2 arms; D: Not specified] |
| 12. Type of endpoint(s) considered | A/ B/ C/ D/ E [A: Binary endpoint; B: Continuous endpoint; C: Survival endpoint; D: Others; E: Not specified] |
| **Statistical methodology** | |
| 13. Method(s) used for incorporating non-current controls | [Text] |
| 14. Type of method to incorporate non-concurrent controls | A/ B/ C [A: Frequentist approach; B: Bayesian approach; C: Hybrid approach] |
| 15. Is the method downweighting the non-concurrent controls depending on the differences between concurrent and non-concurrent controls? | Yes/ No |
| 16. Modelling-based approach: Is the method modelling time? If yes, | Yes/ No |
| 16a. Does it assume any distribution of time trends? | Yes/ No |
| 16b. How are time trends incorporated into the model/ method? | [Text] |
| 17. Can the method adjust for covariates? | A/ B/ C [A: Yes; B: No; C: Not specified] |
| 18. Was the potential bias discussed? | Yes/ No |
| 19. Are interim analyses also covered? | Yes/ No |
| **Simulation/ Case Studies** | |
| 20. Were any simulations or case studies covered? If yes, | Yes/ No |
| 21. What study design was used for the simulations (e.g. allocation ratio, treatment arms, type of endpoint)? | [Text] |
| 22. Was the data-generating model described? If yes, | Yes/ No |
| 22a. Which distribution was used to simulate the responses? | [Text] |
| 23. Were the simulations based on a real dataset? If yes, | Yes/ No |
| 23a. Was a real data set used to decide on the parameter values? | Yes/ No |
| 24. How was the performance of the method evaluated? | [Text] |
| 25. Were scenarios under the null hypothesis considered? | Yes/ No |
| 26. Were scenarios under the alternative hypothesis considered? | Yes/ No |
| 27. Were time trend patterns simulated? If yes, | Yes/ No |
| 27a. Which time trend patterns were used? | [Text] |
| 28. Were the simulations performed to compare the methods to another? | Yes/ No |
| **Software** | |
| 29. What package and/ or software was used? | [Text] |
| 30. Is the software or code available? | Yes/ No |
| **Limitations and Conclusion** | |
| 31. What are limitations of the study and what is the general conclusion? | [Text] |

Table 2: List of included publications

| Title | First author | Journal |
|---|---|---|
| Minimizing control group allocation in randomized trials using dynamic borrowing of external control data - An application to second line therapy for non-small cell lung cancer | Dron L | Contemp Clin Trials |
| Utilizing shared internal control arms and historical information in small-sized platform clinical trials | Jiao F | J Biopharm Stat |
| Borrowing from Historical Control Data in Cancer Drug Development: A Cautionary Tale and Practical Guidelines | Lewis CJ | Stat Biopharm Res |
| A Comparison Between a Meta-analytic Approach and Power Prior Approach to Using Historical Control Information in Clinical Trials With Binary Endpoints | Isogawa N | Ther Innov Regul Sci |
| Incorporating individual historical controls and aggregate treatment effect estimates into a Bayesian survival trial: a simulation study | Brard C | BMC Med Res Methodol |
| Design of randomized controlled confirmatory trials using historical control data to augment sample size for concurrent controls | Yuan J | J Biopharm Stat |
| A practical Bayesian adaptive design incorporating data from historical controls | Psioda MA | Stat Med |
| Bayesian selective response-adaptive design using the historical control | Kim MO | Stat Med |
| Use of a historical control group in a noninferiority trial assessing a new antibacterial treatment: A case study and discussion of practical implementation aspects | Dejardin D | Pharm Stat |
| Covariate-adjusted borrowing of historical control data in randomized clinical trials | Han B | Pharm Stat |
| Robust meta-analytic-predictive priors in clinical trials with historical control information | Schmidli H | Biometrics |
| Use of historical control data for assessing treatment effects in clinical trials | Viele K | Pharm Stat |
| Using historical control information for the design and analysis of clinical trials with overdispersed count data | Gsteiger S | Stat Med |
| Adaptive adjustment of the randomization ratio using historical control data | Hobbs BP | Clin Trials |
| The inclusion of historical control data may reduce the power of a confirmatory study | Cuffe RL | Stat Med |
| Incorporating historical control data in planning phase II clinical trials | Thall PF | Stat Med |
| Statistical considerations of phase 3 umbrella trials allowing adding one treatment arm mid-trial | Ren Y | Contemp Clin Trials |
| The Use of External Control Data for Predictions and Futility Interim Analyses in Clinical Trials | Ventz S | Neuro-Oncology |
| The Use of External Controls in FDA Regulatory Decision Making | Jahanshahi M | Ther Innov Regul Sci |
| The use of external controls: To what extent can it concurrently be recommended? | Burger HU | Pharm Stat |
| Bayesian semiparametric meta-analytic-predictive prior for historical control borrowing in clinical trials | Hupf B | Stat Med |
| A novel equivalence probability weighted power prior for using historical control data in an adaptive clinical trial design: A comparison to standard methods | Bennett M | Pharm Stat |
| A roadmap to using historical controls in clinical trials - by Drug Information Association Adaptive Design Scientific Working Group (DIA-ADSWG) | Ghadessi M | Orphanet J Rare Dis |
| Historical Controls in Randomized Clinical Trials: Opportunities and Challenges | Hall KT | Clin Pharmacol Ther |
| Including non-concurrent control patients in the analysis of platform trials: is it worth it? | Lee KM | BMC Med Res Methodol |
| Reducing Patient Burden in Clinical Trials Through the Use of Historical Controls: Appropriate Selection of Historical Data to Minimize Risk of Bias | Lim J | Ther Innov Regul Sci |
| Propensity score-integrated composite likelihood approach for augmenting the control arm of a randomized controlled trial by incorporating real-world data | Chen WC | J Biopharm Stat |
| Critical appraisal of Bayesian dynamic borrowing from an imperfectly commensurate historical control | Harun N | Pharm Stat |
| An efficient Bayesian platform trial design for borrowing adaptively from historical control data in lymphoma | Normington J | Contemp Clin Trials |
| Beyond Randomized Clinical Trials: Use of External Controls | Schmidli H | Clin Pharmacol Ther |
| Clustered allocation as a way of understanding historical controls: Components of variation and regulatory considerations | Collignon O | Stat Methods Med Res |
| Bayesian leveraging of historical control data for a clinical trial with time-to-event endpoint | Roychoudhury S | Stat Med |
| A note on the power prior | Neuenschwander B | Stat Med |
| Power Prior Distributions for Regression Models | Ibrahim JG | Stat Sci |
| Modified power prior with multiple historical trials for binary endpoints | Banbeta A | Stat Med |
| Elastic priors to dynamically borrow information from historical data in clinical trials | Jiang L | Biometrics |
| The combination of randomized and historical controls in clinical trials | Pocock | J Chron Dis |
| Hierarchical Commensurate and Power Prior Models for Adaptive Incorporation of Historical Information in Clinical Trials | Hobbs BP | Biometrics |
| Elastic meta-analytic-predictive prior for dynamically borrowing information from historical data with application to biosimilar clinical trials | Zhang W | Contemp Clin Trials |
| Summarizing historical information on controls in clinical trials | Neuenschwander B | Clin Trials |
| A Bayesian model with application for adaptive platform trials having temporal changes | Wang C | Biometrics |
| The Bayesian Time Machine: Accounting for Temporal Drift in Multi-arm Platform Trials | Saville B. R. | Clin Trials |
| Adaptive power priors with empirical Bayes for clinical trials | Gravestock | Pharm Stat |

# 2 Additional information regarding the guideline review

Next, we describe the protocol followed for the guideline review. We also present the extraction form for this review and list the reviewed guidelines.

## 2.1 Protocol for the guideline review

This systematic review was performed in accordance with the PRISMA reporting guideline's extension for scoping reviews [7]. The database for our systematic review of guidelines was based on all guidelines available for download on 20/05/2021. We accessed all scientific guidelines from the database of the European Medicine Agency[1] (EMA) with the following filters

- "Topic=Scientific guidelines",
- "Categories=Human",
- "Type of content=Documents" and
- "Include Documents=Yes"

while we used the following filters for the database of the U.S. Food and Drug Administration[2] (FDA)

- "Product=Drugs" and
- "Product=Biologics"

Then, we used the advanced search function of Adobe Acrobat Pro 2020 to search in all the PDF documents for the following keywords:

```
non-concurrent control(s)
concurrent control(s)
historical control(s)
shared control(s)
historical borrowing
external control(s)
master protocol(s)
bayesian method(s)
```

---

[1] https://www.ema.europa.eu/en/search/search
[2] https://www.fda.gov/regulatory-information/search-fda-guidance-document

Duplicates and older draft versions were excluded after the keyword search. Afterwards, we screened the identified documents and included

- guideline documents,
- Questions and Answers (QnAs),
- qualification opinion and
- reflection papers

from EMA, ICH or FDA for text extraction. We excluded documents

- in which external/non-concurrent controls were not discussed in the context of an inclusion into the primary analysis (e.g. the use of external/non-concurrent controls were just mentioned in the context of sample size planning),
- in which one of the keywords and hence, the use of external/non-concurrent controls was only mentioned without further recommendation or description (e.g. mentioned only in the title of a reference)
- in which the use of external/non-concurrent controls was only mentioned in a non-clinical or preclinical setting
- or in the context of (secondary) safety data analyses or meta-analyses
- in which the use of external/non-concurrent controls was discussed in a medical device context.

Following the inclusion and exclusion of documents, two independent reviewers used a standardized data extraction form (see Table 3) to extract the relevant information. Where possible, pre-specified categories were defined for each item in the extraction form. General information of the guidelines such as the year of the guideline or the type of document (e.g. guideline, reflection paper) were extracted. We further documented whether the guideline discussed the use of external/non-concurrent data in an early or late phase and whether the guideline was focused on methodological or clinical aspects. Concerning the use of external/non-concurrent controls, details on the specific circumstances were extracted in which the use was recommended or deemed acceptable, or unacceptable, the concerns that were raised as well as the requirements for the use. Furthermore, the methods mentioned for the incorporation of external/non-concurrent data and the type of inferential question addressed were collected. We specifically identified whether the use of non-concurrent controls, or the joint use of external and concurrent controls in platform trials was discussed in the guideline. Adjudication was performed by a third reviewer in case of discrepancies.

After the text extraction, the number of guidelines in each pre-specified category was determined and free-text fields were summarized in listings.

Overall, 1527 documents were downloaded from the EMA and FDA database (download date: 20/05/2021): 176 documents from the EMA database and 1351 documents from the FDA website. After searching all the documents for the above keywords, 232 documents were identified. In total, 97 documents were included for the screening after exclusion of duplicates and old drafts. Based on the above inclusion and exclusion criteria, 60 guidelines were excluded, resulting in 37 guidelines for the final review and data extraction (for a full list of included guidelines see Table 4).

Table 3: Extraction form for guideline review

| Question | Possible options or [Data type] |
|---|---|
| Title of the document | [Text] |
| 1. Year of the document | [integer] [Note: if unknown, leave blank] |
| 2. Type of document | Guideline/Reflection paper/QnA/Other |
| 3. Source of the guideline | FDA/ICH/EMA |
| 4. Does the document focus on the use of historical/external control in early or late phase trials? | Early/Late/Both/Not mentioned |
| 5. General methodological or clinical indication specific document? | Methodological/Clinical/Both/Unclear |
| 6. Use of external controls under specific circumstances recommended or deemed acceptable? | 3/2/1/0 [ordinal: 3=Recommended; 2= Not recommended but acceptable; 1= Not acceptable, 0=Unclear] |
| 7. If yes, what are the specific circumstances mentioned in which the use might be acceptable | [Note: if not recommended in 6, leave subquestions blank, "No" here means, that the circumstance is not specifically mentioned in the guideline] |
| Rare disease/event | Yes/No |
| Pediatric | Yes/No |
| Unmet medical need | Yes/No |
| High mortality | Yes/No |
| Long treatment period before endpoint can be measured | Yes/No |
| Indication specific | Yes/No |
| If indication specific yes, which indication: | [Text] [Note: if unknown, leave blank] |
| Large treatment effect | Yes/No |
| No time trend in disease population/management | Yes/No |
| Homogenous treatment effect | Yes/No |
| Ethical concerns regarding assignment to placebo/control | Yes/No |
| Predictable disease course/Natural history well defined | Yes/No |
| Predictable mortality | Yes/No |
| Objective endpoint | Yes/No |
| Drug effect self-evident | Yes/No |
| Feasibility of randomized trial | Yes/No |
| Other | [Text] [Note: if unknown, leave blank] |
| 8. Concerns and/or requirements raised with the use of non-concurrent control data? | |
| Bias | Yes/No |
| Comparability | Yes/No |
| Data/Trial integrity | Yes/No |
| Indication specific concern or requirement | Yes/No |
| If indication specific yes, which indication specific concern: | [Text] [Note: if unknown, leave blank] |
| Yes, but no further details | Yes/No |
| Differences in measurements | Yes/No |
| Change in SOC | Yes/No |
| Selection bias | Yes/No |
| High quality of data | Yes/No |
| Other | [Text] [Note: if unknown, leave blank] |
| 9. Methods mentioned for the incorporation | |
| Matching approach | Yes/No |
| Bayesian method (Power Prior, MAP, …) | Yes/No |
| Meta-analysis | Yes/No |
| Full pooling | Yes/No |
| Threshold crossing, boundary for hypothesis testing | Yes/No |
| Modelling approach (e.g. regression model) | Yes/No |
| Other | [Text] [Note: if unknown, leave blank] |
| 10. Use of Bayesian methods supported? | 2/1/0 [ordinal: 2=Discussed and supported; 1=Discussed and not supported; 0=Not discussed] |
| 11. Type of inferential question addressed in guideline: | |
| Non-inferiority | Yes/No |
| Superiority | Yes/No |
| Equivalence | Yes/No |
| not specified | Yes/No |
| 12. Is the use of non-concurrent controls platform trials discussed? | Yes/No |
| 13. Is specifically the joint use of external and concurrent controls discussed? | Yes/No |

Table 4: List of included guidelines

| Guideline | Year | Source |
|---|---|---|
| ICH guideline E8 (R1) on general considerations for clinical studies | 2019 | ICH |
| Guidance for Industry -E 10 Choice of Control Group and Related Issues in Clinical Trials | 2001 | ICH |
| Guidance for Industry - E11(R1) Addendum: Clinical Investigation of Medicinal Products in the Pediatric Population | 2018 | ICH |
| Guidance for Industry - Neglected Tropical Diseases of the Developing World: Developing Drugs for Treatment or Prevention | 2014 | FDA |
| Guidance for Industry - Non-Inferiority Clinical Trials to Establish Effectiveness | 2016 | FDA |
| Inborn Errors of Metabolism That Use Dietary Management: Considerations for Optimizing and Standardizing Diet in Clinical Trials for Drug Product Development | 2018 | FDA |
| Guidance for Industry - Rare Diseases: Common Issues in Drug Development - Draft Guidance | 2019 | FDA |
| Demonstrating Substantial Evidence of Effectiveness for Human Drug and Biological Products | 2019 | FDA |
| Interacting with the FDA on Complex Innovative Trial Designs for Drugs and Biological Products | 2020 | FDA |
| GUIDELINE FOR THE FORMAT AND CONTENT OF THE CLINICAL AND STATISTICAL SECTIONS OF AN APPLICATION | 1988 | FDA |
| ICH Topic E3 Structure and Content of Clinical Study Reports | 1996 | ICH |
| Guidance for Industry - Influenza: Developing Drugs for Treatment and/or Prophylaxis | 2011 | FDA |
| Guidance for Industry - Time and Extent Applications for Nonprescription Drug Products | 2011 | FDA |
| Guidance for Industry - Expedited Programs for Serious Conditions – Drugs and Biologics | 2014 | FDA |
| Guidance for Industry - Chronic Hepatitis C Virus Infection: Developing Direct-Acting Antiviral Drugs for Treatment | 2017 | FDA |
| Amyotrophic Lateral Sclerosis: Developing Drugs for Treatment | 2019 | FDA |
| Acute Myeloid Leukemia: Developing Drugs and Biological Products for Treatment | 2020 | FDA |
| Guidance for Industry - COVID-19: Master Protocols Evaluating Drugs and Biological Products for Treatment or Prevention | 2021 | FDA |
| Postapproval Pregnancy Safety Studies | 2019 | FDA |
| Guidance for Industry -Antibacterial Therapies for Patients With an Unmet Medical Need for the Treatment of Serious Bacterial Diseases | 2017 | FDA |
| Guidance for Industry - Duchenne Muscular Dystrophy and Related Dystrophinopathies: Developing Drugs for Treatment | 2018 | FDA |
| Guidance for Industry -Clinical Trial Endpoints for the Approval of Cancer Drugs and Biologics | 2018 | FDA |
| Adaptive Designs for Clinical Trials of Drugs and Biologics | 2019 | FDA |
| Reflection paper on the use of extrapolation in the development of medicines for paediatrics | 2018 | EMA |
| Guideline on quality, non-clinical and clinical requirements for investigational advanced therapy medicinal products in clinical trials | 2019 | EMA |
| GUIDELINE ON THE EVALUATION OF ANTICANCER MEDICINAL PRODUCTS IN MAN | 2019 | EMA |
| Guideline on the evaluation of medicinal products indicated for treatment of bacterial infections | 2011 | EMA |
| Reflection paper on the regulatory requirements for vaccines intended to provide protection against variant strain(s) of SARS-CoV-2 | 2021 | EMA |
| Qualification opinion on Cellular therapy module of the European Society for Blood & Marrow Transplantation (EBMT) Registry | 2019 | EMA |
| Expedited Programs for Regenerative Medicine Therapies for Serious Conditions | 2019 | FDA |
| Human Gene Therapy for Neurodegenerative Diseases | 2021 | FDA |
| Safety, Efficacy, and Pharmacokinetic Studies to Support Marketing of Immune Globulin Intravenous (Human) as Replacement Therapy for Primary Humoral Immunodeficiency | 2008 | FDA |
| Considerations for Allogeneic Pancreatic Islet Cell Products | 2009 | FDA |
| Preparation of IDEs and INDs for Products Intended to Repair or Replace Knee Cartilage | 2011 | FDA |
| Clinical Considerations for Therapeutic Cancer Vaccines | 2011 | FDA |
| Human Gene Therapy for Rare Diseases | 2020 | FDA |
| FDARA Implementation Guidance for Pediatric Studies of Molecularly Targeted Oncology Drugs: Amendments to Sec. 505B of the FD&C Act | 2021 | FDA |



\clearpage

\end{document}